\documentclass{emulateapj}
\usepackage{epsf}
\usepackage{apjfonts}
\usepackage{xspace}
\usepackage{amsmath}
\usepackage{framed} 

\usepackage{graphicx}
\usepackage[space]{grffile}
\usepackage{latexsym}
\usepackage{amsfonts,amsmath,amssymb}
\usepackage{url}
\usepackage[utf8]{inputenc}
\usepackage{fancyref}
\usepackage{hyperref}
\hypersetup{colorlinks=true,pdfborder={0 0 0},}
\usepackage{textcomp}
\usepackage{longtable}
\usepackage{multirow,booktabs}
   \usepackage{graphicx}

\def\Mpch{h^{-1}\dim{Mpc}}

\def\HII{\mathrm{HII}}

\def\HI{{\rm H\,I}}
\def\HII{{\rm H\,II}}
\def\HeII{{\rm He\,II}}

\def\dim#1{\mbox{\,#1}}

\def\hide#1{}

\begin{document}

\title{Cosmic Reionization On Computers III. The Clumping Factor}

\author{Alexander A.\ Kaurov\altaffilmark{1} and Nickolay Y.\ Gnedin\altaffilmark{2,1,3}}
\altaffiltext{1}{Department of Astronomy \& Astrophysics, The
  University of Chicago, Chicago, IL 60637 USA; kaurov@uchicago.edu}
\altaffiltext{2}{Particle Astrophysics Center, 
Fermi National Accelerator Laboratory, Batavia, IL 60510, USA; gnedin@fnal.gov}
\altaffiltext{3}{Kavli Institute for Cosmological Physics and Enrico
  Fermi Institute, The University of Chicago, Chicago, IL 60637 USA} 

\begin{abstract}
We use fully self-consistent numerical simulations of cosmic reionization, completed under the Cosmic Reionization On Computers (CROC)  project, to explore how well the recombinations in the ionized IGM can be quantified by the effective ``clumping factor''. The density distribution in the simulations (and, presumably, in a real universe) is highly inhomogeneous and more-or-less smoothly varying in space. However, even in highly complex and dynamic environments the concept of the IGM remains reasonably well-defined; the largest ambiguity comes from the unvirialized regions around galaxies that are over-ionized by the local enhancement in the radiation field (``proximity zones''). That ambiguity precludes computing the IGM clumping factor to better than about 20\%. We also discuss a ``local clumping factor'', defined over a particular spatial scale, and quantify its scatter on a given scale and its variation as a function of scale.
\end{abstract}


\section{Introduction}
\label{sec:intro}

Theoretical study of the epoch of reionization is becoming an increasingly important area of research, as the computational capabilities advance to the point when it becomes possible to run numerical simulations with self-consistent radiation transfer and star formation in cosmological size boxes \citep{newrei:snr14,newrei:nrs14,newrei:hdp14,Gnedin2014a,Gnedin2014b}.

Even though it is common to discuss the sources of ionization radiation, the competing process, recombination, is not less important in the context of reionization. There are two distinct ways of thinking about recombinations during the epoch of reionization. The first one treats the gas density, temperature, and ionized fraction distributions in the universe as continuous fields, with the recombination rate having appropriate values everywhere in space; this is the approach used in advanced simulations. 

The second approach is usually adopted in analytical models and is sometimes used in interpreting observations. It focuses on ionizing photons which, first, escape a host galaxy with some probability, then freely travel through the ionized Intergalactic Medium (IGM), and finally are absorbed either at ionization fronts between the ionized and neutral regions in the IGM or at Lyman Limit systems (LLS). Recombination inside the host galaxy are treated as a reduction in the source emission and are parametrized with the escape fraction \citep{Gnedin2008a, Kuhlen2012}; recombination in the ionized IGM are quantified by the so-called ``clumping factor'', and absorption by LLS are accounted for by an upper limit on the photon mean free path \citep{Songaila2010, Kaurov2013a}. Even though spontaneous recombination is the same physical process no matter where it occurs, three unrelated quantities -- the escape fraction, the clumping factor, and the maximum mean free path -- are used to describe it. 

Our goal in this paper is to examine whether this ``three-regime'' approach is well defined in the fully self-consistent numerical simulations (which we consider as a plausible model of the real universe). Since the recombination rate varies smoothly in space, such a discrete separation of a continuous function into three distinct regimes would only make sense if the actual distribution of recombination rates is, in some particular way, tri-modal and a clear separation can be made between recombinations in the ISM (the escape fraction), IGM (the clumping factor), and in the LLS (the maximum mean free path).

In \S\ref{sec:phasediag} we describe a useful phase diagram, which helps to define these three regimes in a well motivated (rather than based on some arbitrary density or ionized fraction thresholds) way. Hence, if carefully made, such a separation of cosmic recombinations into distinct regimes can indeed be reasonable, and the quantities such as the IGM clumping factor can be meaningfully defined.

Then, in \S\ref{sec:clumpingfactor}, we focus on the ionized IGM and its clumpiness, since it occupies the majority of volume and defines the morphology of reionization. Due to its quadratic dependence on the density, the recombination rate inside a given volume $V$ explicitly depends on the actual density distribution inside the volume. It is convenient to use the clumping factor 
\begin{equation}
C=\langle n_i^2\rangle / \langle n_i \rangle ^2, 
\label{eq:cf}
\end{equation}
to factorize out that dependence and to express the recombination rate through the mean density inside the volume $V$,
\begin{equation}
R = \int_V n_i^2 \alpha dV \approx \alpha C  \langle n_i \rangle_V ^2 V,
\end{equation}
where $n_i$ is the number density of ionized hydrogen (we assume that the number of electrons is proportional to the number of ionized hydrogen, which is the case before \HeII\ reionization), $R$ is the recombination rate inside the volume $V$, and $\alpha(T)$ is the recombination coefficient. 

However, the lack of a common definition of the ionized IGM leads to problems with comparing the clumping factor between different numerical studies \citep{Gnedin1997a,Iliev2005,
Kohler2007,McQuinn2007,Trac2007,
Pawlik2009,Raivcevic2011,Finlator2012,
Shull2012,Emberson2013,So2014,Jeeson-Daniel2014} and analytical models \citep{Sobacchi2014,Kaurov2014}. We reexamine the value of the clumping factor with \cite{Gnedin2014a} simulation described in \S\ref{sec:Simulation} and discuss its definition. In addition, we study its spatial inhomogeneity in \S\ref{sec:localclumpingfactor} and its correlation with density in \S\ref{subsec:localclump}. 

\section{Simulation}
\label{sec:Simulation}

As a physically plausible model of reionization we use numerical simulations from the Cosmic Reionization On Computers (CROC) project \citep{Gnedin2014a, Gnedin2014b}. 

These simulations are suitable for our purposes for several reasons. CROC simulations include the whole range of physical processes required in order to model reionization fully self-consistently\footnote{We purposedly distinguish terms ``self-consistent'' and ``from the first principles''. CROC simulations do include free parameters in the underlying physical model (gas depletion time due to star formation, delayed cooling time scale, effective emissivity of ionizing photons at the simulation resolution limit, etc), hence they are not ``from the first principles'' simulations. We call them ``self-consistent'', though, in a sense that all modeled physical processes are followed with the same spatial and temporal resolution, co-evolving and affecting each other as the simulation proceeds. In that sense the term ``self-consistent'' distinguishes CROC simulations from numerical models where radiative transfer is done in post-processing, or which track hydrodynamics and radiative transfer with two separate numerical schemes with widely divergent resolutions.}, from gas dynamics to fully coupled 3D radiative transfer, star formation, and stellar feedback. They match the existing observational constraints on the evolution of galaxy luminosity functions and on the full distribution function of Gunn-Peterson absorption in the spectra of high redshift quasars. Using the DC model formalism of \citet{gkr11}, the simulations account for the cosmic variance between several independent realizations. Finally, using Adaptive Mesh Refinement, CROC simulations achieve spatial resolution of $125\dim{pc}$ in simulation volumes of up to $40h^{-1}$ comoving Mpc.

Such high spatial resolution allows us to consider everything outside galaxies well resolved, and therefore, such quantities as clumping factor can be computed directly with no prior assumptions. The internal structure of galaxies is not well resolved with the spatial resolution of $\sim100\dim{pc}$; therefore, we do not consider clumping of the ISM in this paper.

In this paper we use three $40h^{-1}\dim{Mpc}$ simulations with different DC modes (runs B40.sf1.uv2.bw10.A-C from \citet{Gnedin2014a}) as our fiducial set, with all presented quantities averaged over these three runs. In addition, we use two $20h^{-1}\dim{Mpc}$ simulations: the ``medium resolution'' one (run B20.sf1.uv2.bw10.B from \citet{Gnedin2014a}) that matches our fiducial set in spatial and mass resolution and a higher resolution simulation (run B20HR.sf1.uv2.bw10.B from \citet{Gnedin2014a}) that we use to test numerical convergence.

\section{Phase diagram}
\label{sec:phasediag}

In this section we describe our main tool (a kind of a ``phase diagram''), which we use in the subsequent sections to classify the cosmic gas distribution into ISM, IGM, and LLS. Specifically, we plot the mass weighted distribution of all cells from a simulation in two dimensions: density, $(1+\delta)$, and \textit{ionization state indicator}, $\varkappa$, which is defined as:
\begin{equation}
      \varkappa \equiv (1+\delta)\;\dfrac{x_\HII^2}{x_\HI},
\label{eq:varkappa0}
\end{equation}
where $x_\HII$ and $x_\HI$ are the fractions of ionized and neutral hydrogen relative to the total abundance of hydrogen.

This definition is motivated by a consideration that in the photoionization equilibrium
\begin{equation}
\label{eq:gamma}
      \Gamma \, n_\HI \, = \, n_\HII^2 \, \alpha(T),
\end{equation}
where $\Gamma$ is ionizing background, and hence
\begin{equation}
\label{eq:varkappa1}
    \varkappa \propto \frac{\Gamma}{\alpha(T)}.
\end{equation}
In particular, in the IGM with modest temperature fluctuations and approximately homogeneous ionizing background, $\varkappa \approx \mbox{const}$.

Thus, the ionization state indicator is a good tracer of general IGM under the assumption of: (a) uniform ionizing background, (b) no collisional ionization \footnote{In case of presence of collisional ionization, the photoionization equilibrium does not hold. Thus, Equation \ref{eq:gamma} should have another term on the left hand side.}, and (c) uniform temperature. Therefore, if the ionizing background is indeed uniform in ionized regions, then we should see a narrow distribution of $\varkappa$.

\begin{figure}[t]
      \begin{center}
      \includegraphics[scale=1.0]{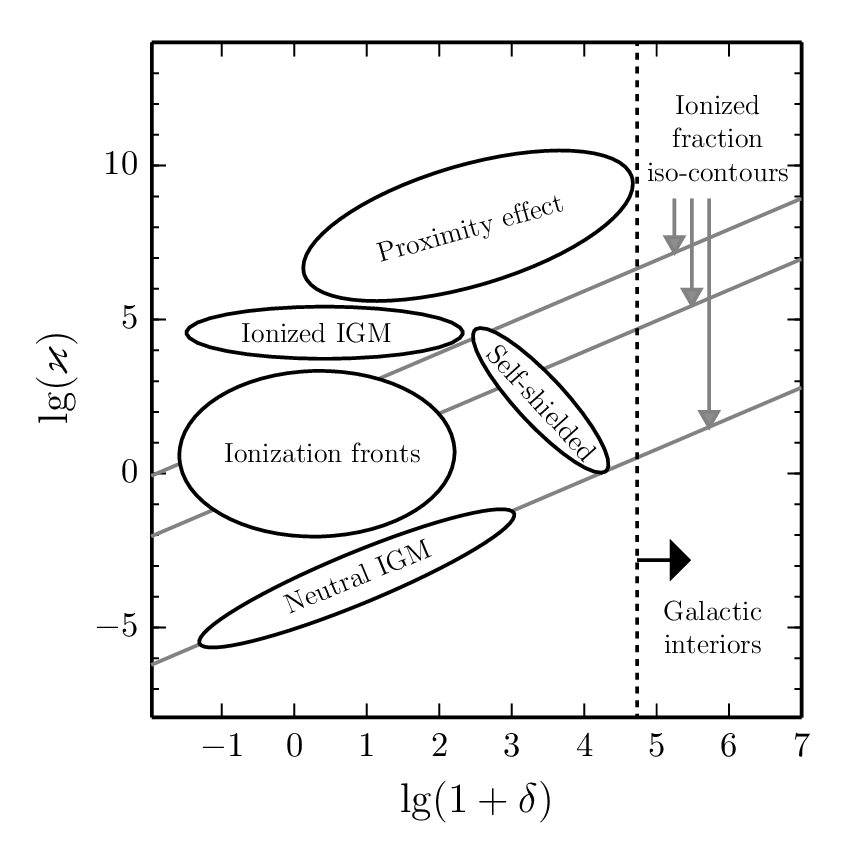}
      \end{center}
    \caption{\label{fig:phaseplot-cartoon}Schematic representation of the phase diagram introduced in \S\ref{sec:phasediag}.}
\end{figure}

A few distinct features, which correspond to different states of gas, emerge from this phase diagram. We label these features in Figure \ref{fig:phaseplot-cartoon} and discuss them individually, as well as the general properties of the phase diagram, in the following subsections. While the overdensity $\delta$ traces the large scale structure and gives an idea about where spatially these regions are located, the variations in the ionization state indicator show where the assumptions (a)-(c) break down.
	
\begin{figure*}[t]
      \begin{center}
      \includegraphics[scale=1.0]{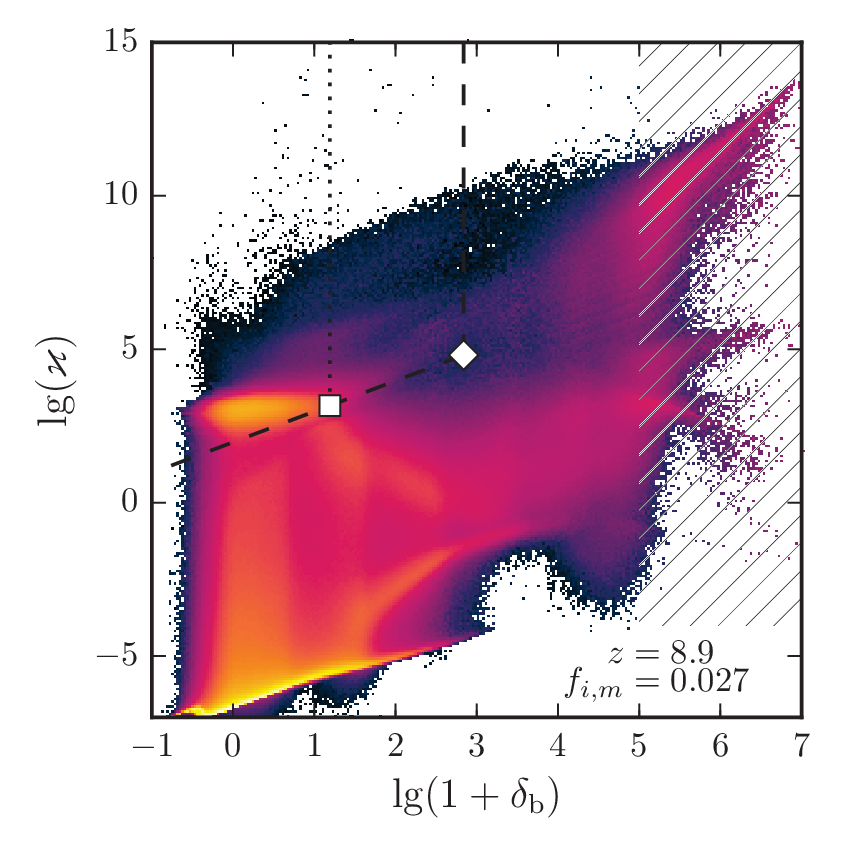}
      \includegraphics[scale=1.0]{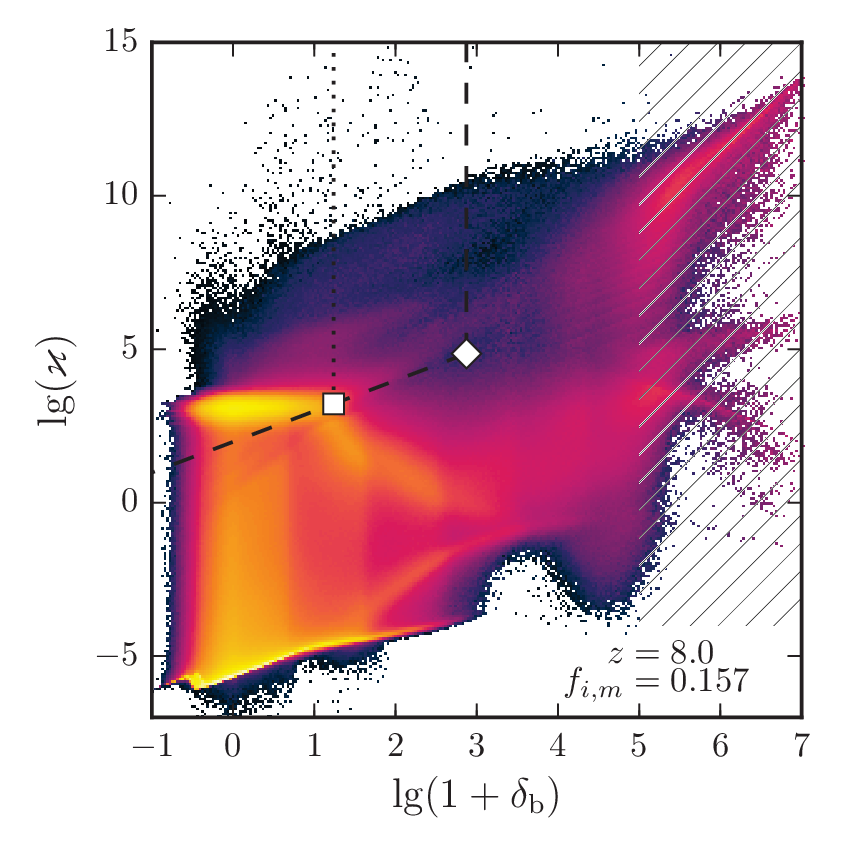}
      \includegraphics[scale=1.0]{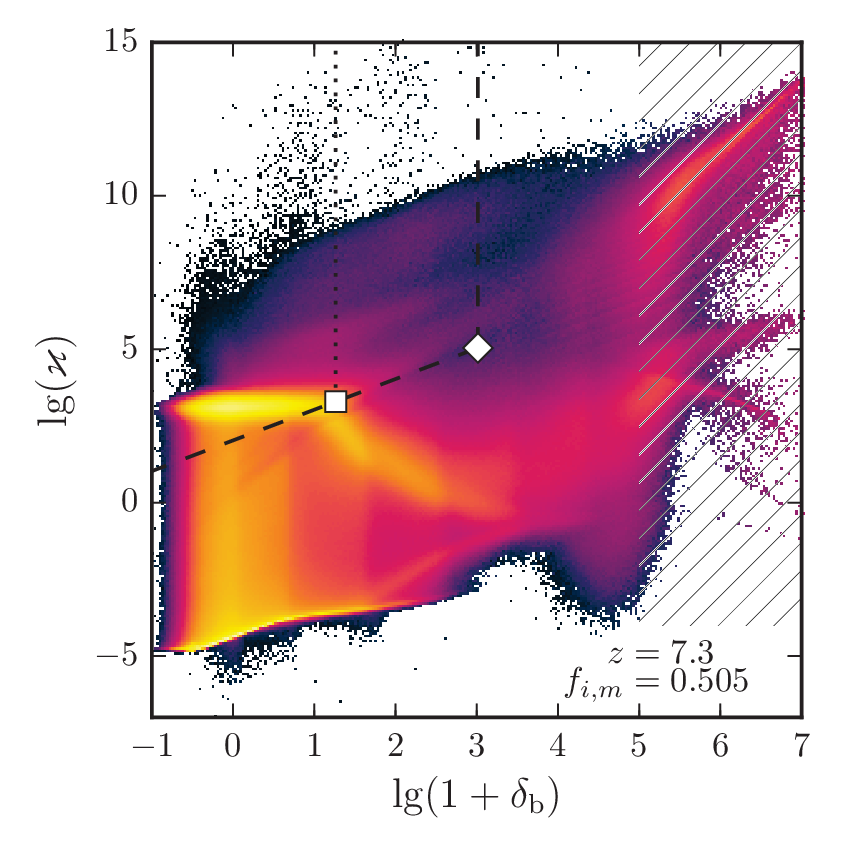}
      \includegraphics[scale=1.0]{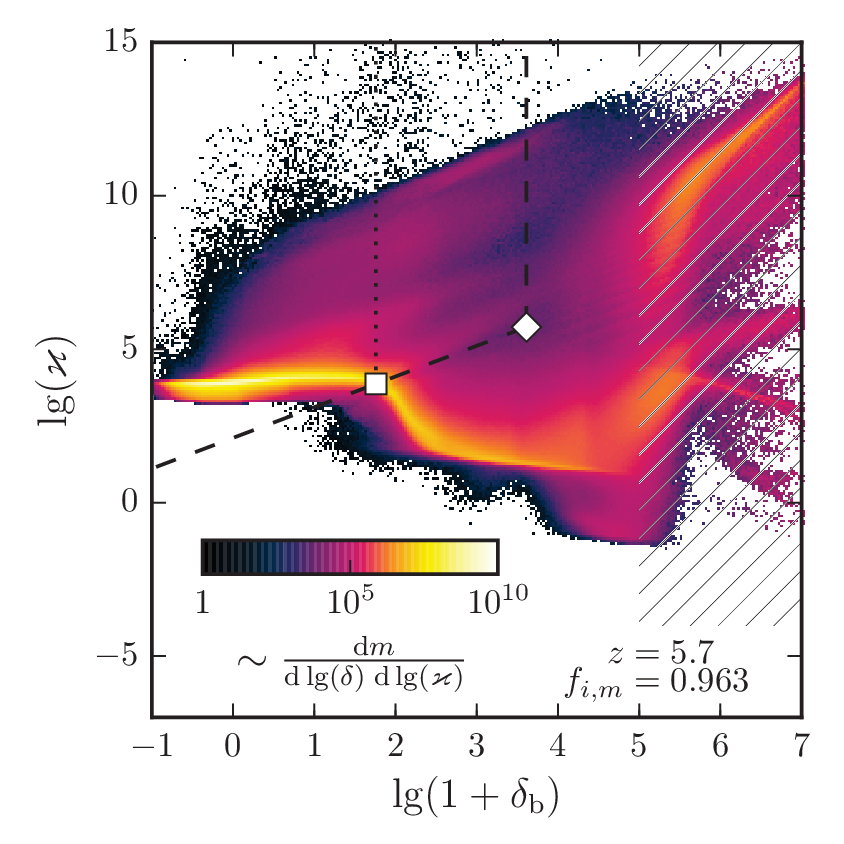}
      \end{center}
    \caption{\label{fig:phaseplot} The $\delta-\varkappa$ phase diagram captured at four redshifts. The mass weighted ionized fraction. $f_{i,m}$, is shown in each panel. Dashed lines show density and ionized fraction thresholds presented in Figure \ref{tab:thresh}. White diamond and square represent the first and the second pivot points we adopt for the definition of the ionized IGM. The hatched area corresponds to the high density regions inside galaxies (IGM), which is not well resolved in the simulation.}
\end{figure*}

\subsection{General properties of the $\delta-\varkappa$ phase diagram}
\label{subsec:deltavarkappa}

To give a better understanding of this type of a phase diagram, we outline its major properties using an optically thin (ionized) Lagrangian volume collapsing into a dense region as an example. Initially it is located in the region labeled as ``Ionized IGM'' in Figure \ref{fig:phaseplot-cartoon}. As the Lagrangian volume gets denser, it moves to the right in the phase diagram. However, the behavior of the $\varkappa$ value depends on the environment.

If the volume contracts slowly enough, it has enough time to recombine and to reach the equilibrium with the ionizing background. In this case the ionization state indicator does not change, and we see a horizontal movement in the phase diagram.

If the same volume collapses rapidly and contracts faster than it recombines, i.e. the ionized fraction does not change, then it moves along the line with the unity slope in $\delta-\varkappa$ space. Therefore, in the phase diagram that region moves up and may escape the ``Ionized IGM'' region into what we label as the ``Proximity effect'' part of the phase diagram.

Another mechanism for a volume to enter ``Proximity'' zone is to approach a source of ionizing radiation. It will locally boost radiation background and, if the medium remains optically thin, $\varkappa$ will also increase.

As the Lagrangian volume continues to collapse, at some point it enters the optically thick regime. The ionization rate inside it drops dramatically; consequently, the ionized fraction decreases, and the ionization state indicator $\varkappa$ decreases too. In the phase diagram it corresponds to the transition to the ``Self-shielded'' region in the phase diagram. 

\subsection{Discussion of individual regions}
\label{subsec:regions}

The actual phase diagrams from the CROC simulations at 4 different redshifts are shown in Figure \ref{fig:phaseplot}. As reionization proceeds, the distribution of gas between various regions in the phase diagram changes.

Ionized and Neutral IGM. Before reionization the IGM has a small uniform ionized fraction left after the recombination epoch. Therefore, at redshifts 8.9, 8.0 and 7.3, when a significant fraction of low density gas is still neutral, we see it distributed along a line with the unity slope. The IGM progressively becomes more and more ionized as redshift approaches 5.7, and occupies different parts of the phase diagram. The value of $\varkappa$ increases, which represents the growth of the ionized fraction. 

Ionization fronts are the intermediate stage between the ionized and neutral IGM. Since a particle spends relatively short time inside an ionization front, only a small fraction of mass is located in that phase at any given time. The fraction of total mass in ionization fronts is order of 5\% at redshifts with 50\% total ionization fraction and lower during other periods (and that is only an upper limit, as ionization fronts are not resolved in CROC simulations).

In individual ionized bubbles the ionization rate may vary significantly from bubble to bubble, due to random variations in the number of sources inside them. As ionized bubbles start to overlap, the ionization rate becomes more and more uniform. This is reflected in the phase diagram in the width of ``Ionized IGM'' region, which becomes much narrower at $z=5.7$ compared to $z=7.3$. The nuances of our specific IGM definition are discussed in \S\ref{sec:clumpingfactor}.

\begin{figure*}[t]
    \begin{center}
    \includegraphics[scale=0.55]{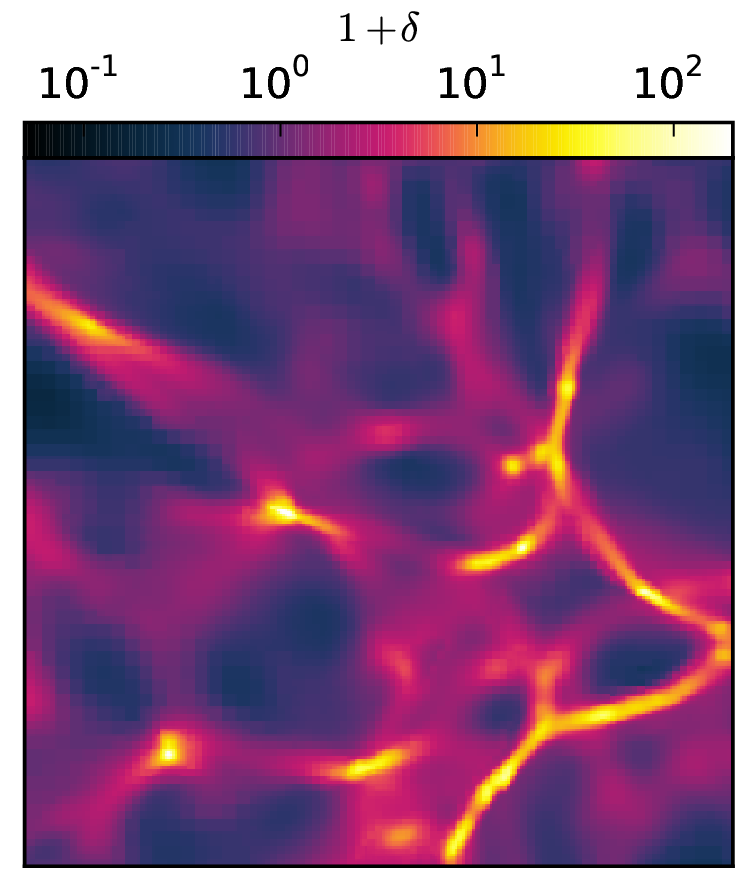}
    \includegraphics[scale=0.55]{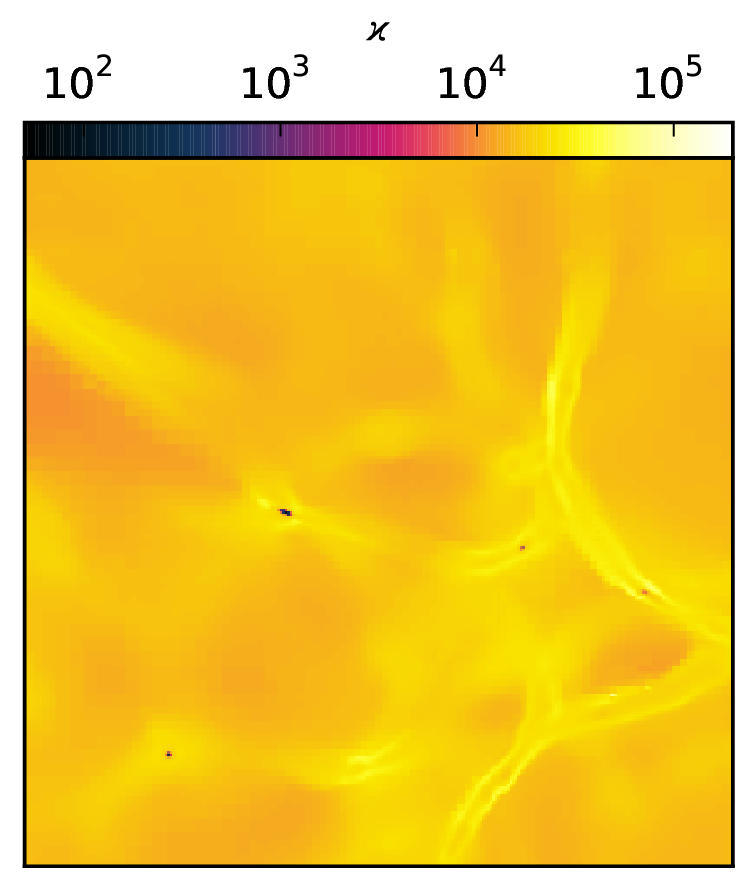}
    \includegraphics[scale=0.55]{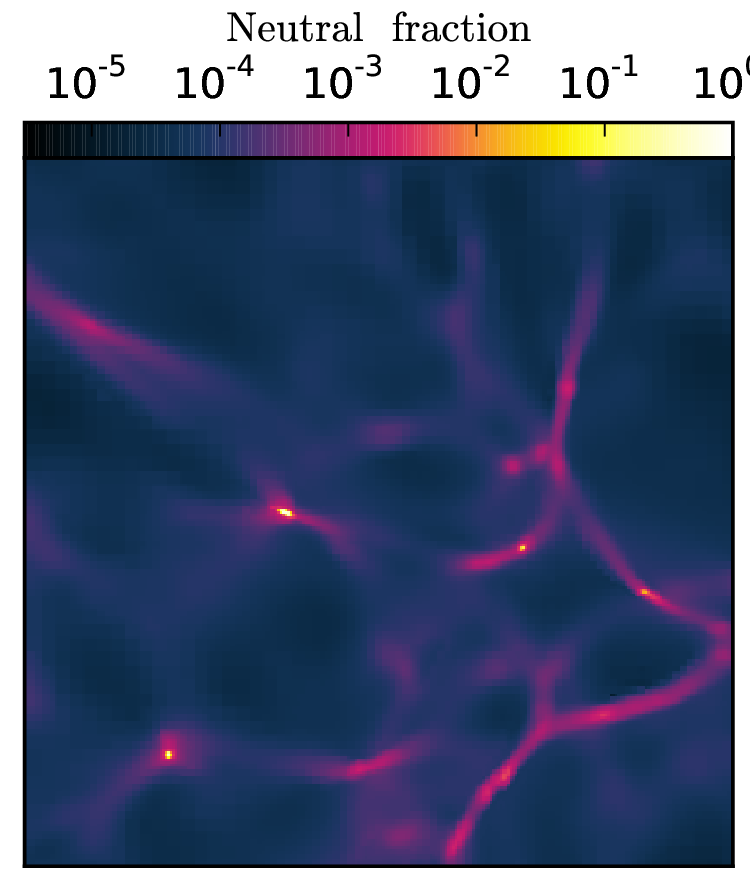}
    \includegraphics[scale=0.55]{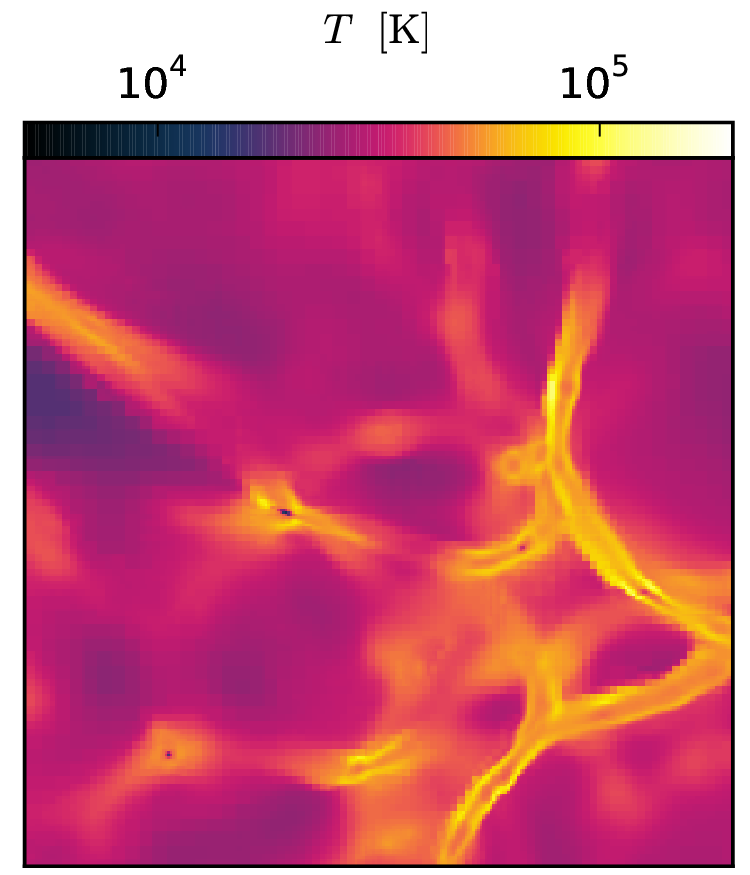}
    \end{center}
    \caption{\label{fig:selfshield}$2\Mpch\times 2\Mpch$ slice at $z=5.7$ showing filament structure. From left panel to right panel: gas density,  ionization state indicator, neutral fraction of hydrogen, and temperature.}
\end{figure*}

\textbf{Self-shielded} regime. At higher densities (marked with a square in Figure \ref{fig:phaseplot}) self-shielding becomes important. Spatial locations with no or very low star formation correspond to the sharp decline in $\varkappa$ from the typical values in the ionized IGM. As reionization proceeds, the ionizing background builds up, and burns deeper into these regions; therefore, the self-shielded region in the diagram moves to the right with time. In Figure \ref{fig:selfshield} self-shielded regions manifest themselves as tiny dark spots in the ionization state indicator or temperature maps or bright spots in the neutral fraction map.

\textbf{Proximity effect.} After reionization, the scatter in $\varkappa$ in the optically thin regions can occur due to the proximity to ionizing sources, large temperature variations, and non-equilibrium effects.

By looking on spatial distribution of $\varkappa$ we can observe a few other effects. In Figure \ref{fig:selfshield} we look at filament structure at redshift 5.7, however the same behavior is typical for any redshift inside ionized regions. Modest increase in $\varkappa$ is observed around filaments due to the rapid contraction of these regions. It leads to the increase of temperature, decrease of recombination coefficient, and, consequently, growth of ionization state indicator, which is inversely proportional to the recombination coefficient. Additionally, in these relatively sparse regions the characteristic contraction timescale may approach recombination time and, consequently, matter will be more ionized than it would be in the ionization equilibrium. So, the shell around filaments shows slight increase of $\varkappa$, but overall that effect is small.

\begin{figure*}[t]
    \begin{center}
    \includegraphics[scale=0.55]{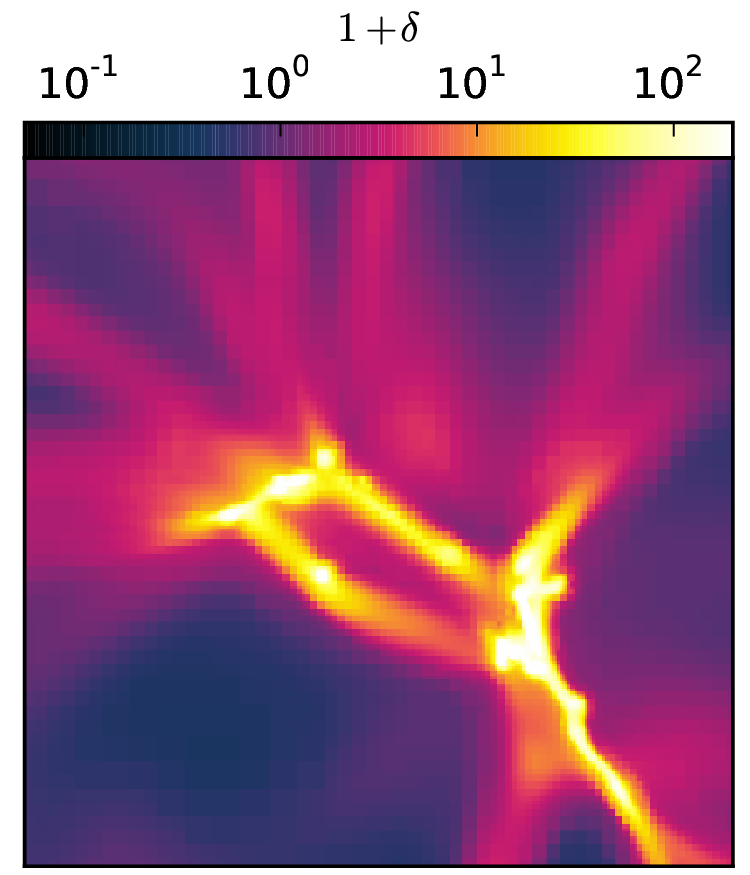}
    \includegraphics[scale=0.55]{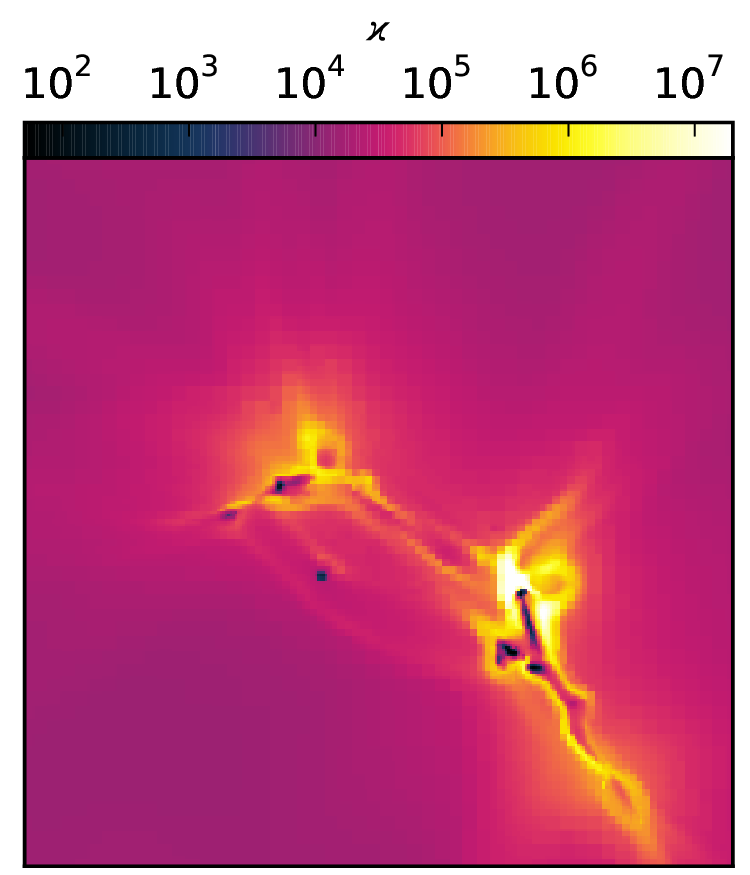}
    \includegraphics[scale=0.55]{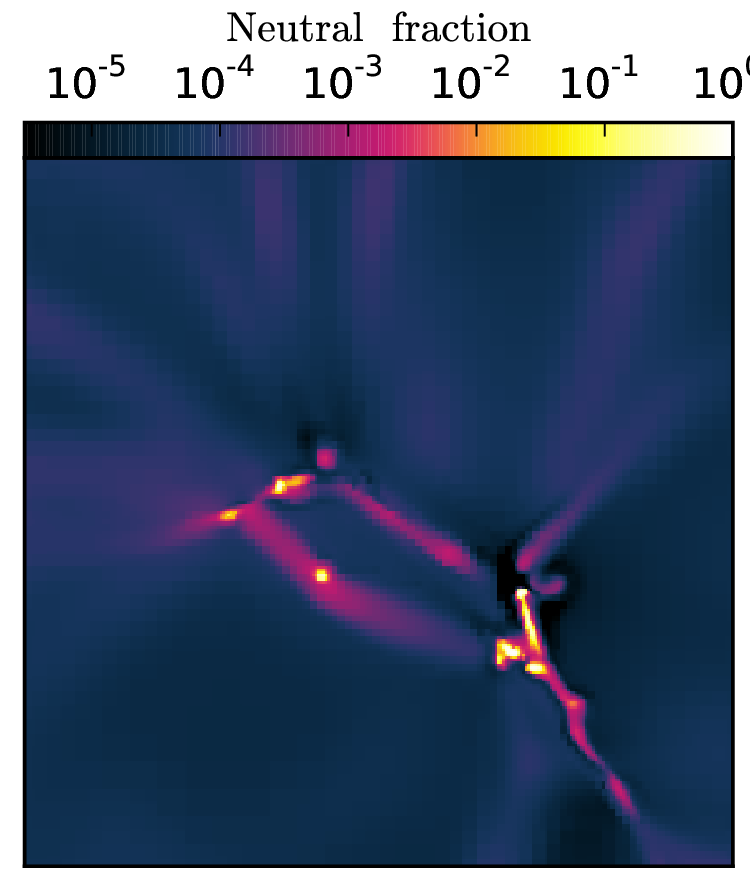}
    \includegraphics[scale=0.55]{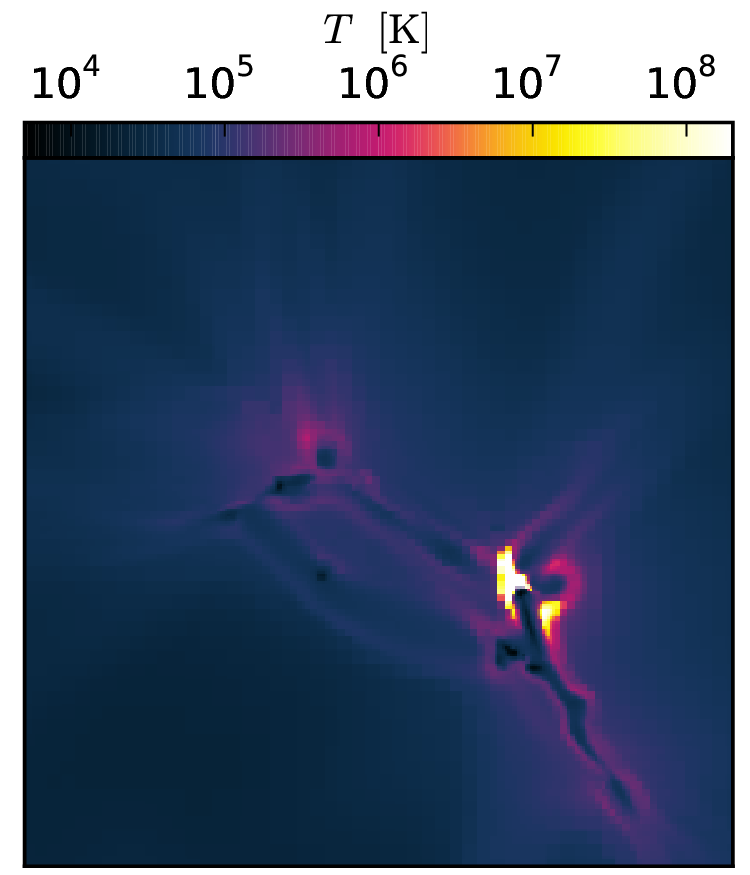}
    \end{center}
    \caption{\label{fig:smerger}$1\Mpch\times 1\Mpch$ slice at $z=5.7$ showing a region with active merging. From left panel to right panel: gas density, ionization state indicator, neutral fraction of hydrogen, and temperature.}
\end{figure*}

In Figure \ref{fig:smerger} we zoom into a merging halo also at reshift 5.7. Regions with active merging stand out because of extreme temperatures. There, beside change in the recombination coefficient, collisional ionization of hydrogen also takes place, driving the neutral fraction away from a pure photoionization equilibrium. These regions are located in top left side of phase diagram -- roughly at $10 < \delta < 10^3$ and $10^5 < \varkappa < 10^{12}$, and contain less than 1\% of the total mass.
    
\textbf{Galactic interiors.} In the simulation that we used, the galactic interiors are not well resolved. Therefore, we do not discuss the features at $\delta\ga10^5$ in the phase diagram - that would require higher resolution simulations of individual galaxies.

\section{IGM clumping factor}
\label{sec:clumpingfactor}

Before calculating the clumping factor of the ionized IGM, one first needs to define what the ionized IGM is. The most common approach of defining the ionized IGM is based on two thresholds: an upper limit on the gas density and a lower limit on the ionization fraction. These cuts exclude neutral dense matter and the ISM from contributing to gas clumping. We adopt the notation $C_{\delta,\,x_{\HII}}$ for the clumping factor calculated over the volume with overdensity below $\delta$ and the ionization fraction above $x_{\HII}$. The choice of these thresholds is somewhat arbitrary, which makes such a definition not well motivated. Here we propose a physically-motivated fix, based on the phase diagram introduced in the previous section.

\begin{figure}
    \begin{center}
    \includegraphics[scale=1.0]{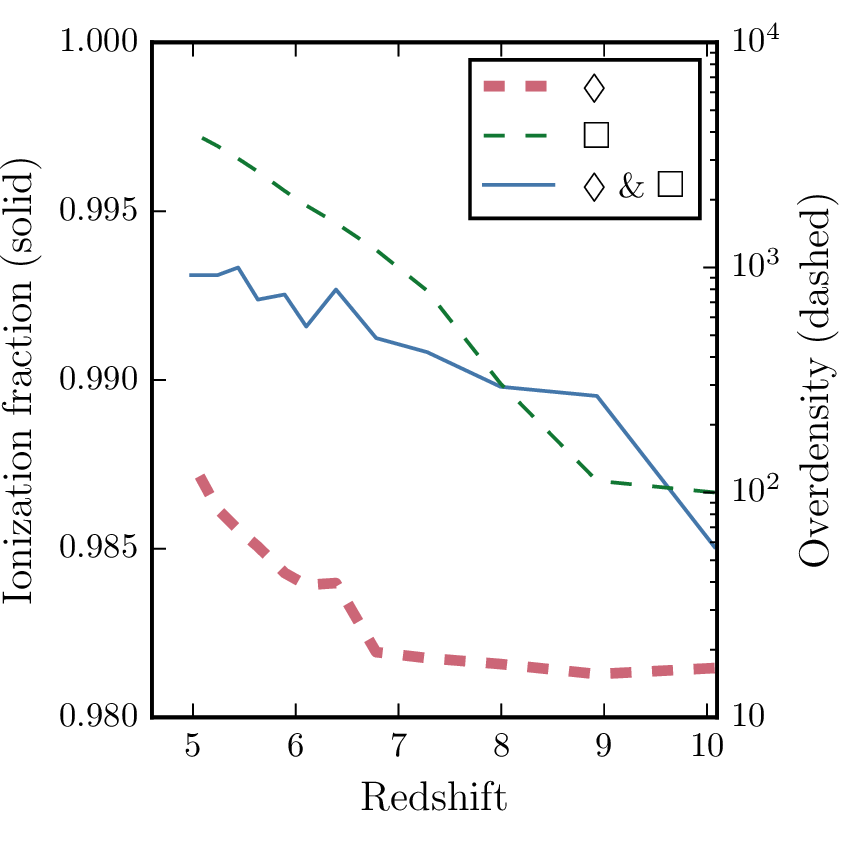}
    \end{center}
    \caption{\label{tab:thresh}Density thresholds used for definition of $C_\Diamond$ (thin green dashed line) and $C_\Box$ (thick red dashed line), along with the ionization threshold used in both definitions (blue solid line).}
\end{figure}

The $\varkappa-\delta$ phase diagram contains a few features, which can be used as pivot points. The most prominent one is the transition between ``Ionized IGM'' and ``Self-shielded'' regions (a square symbol in Figure \ref{fig:phaseplot}). At $z=5.7$ it is easily identifiable along the yellow ridge that contains most of mass in low density regions as a point when the ridge turns down (towards more neutral gas). However, it is less pronounced at high redshifts and also the corresponding overdensity threshold does not fully include ``Proximity'' zone. Instead, we can use another pivot point located at a local minimum between ``Proximity effect'', ``Self-shielded'' and ``Galactic interiors'' zones (a diamond symbol in Figure \ref{fig:phaseplot}). It is prominent at all redshifts, and includes ``Proximity'' zone, and  mathematically well defined, making it much easier to find with a simple algorithm. For each redshifts we determine this point and record corresponding overdensity and ionized fraction (the actual values are plotted in Figure \ref{tab:thresh}). 

At redshifts where both these points are well defined, they always lie at the same value of the ionized fraction. It is not clear whether this is a universal property of the $\varkappa-\delta$ phase diagram or a mere coincidence, but that fact does not appear to be important enough to warrant a targeted study. Hence, using that property, we can define the transition into the ``Self-shielded'' regime at all redshifts, as a local maximum in the phase diagram along the line of constant ionized fraction (dashed line) passing through the diamond pivot point.

The values of the pivot points can be used in defining physically-motivated thresholds in the definition of the ionized IGM, although one can, in principle, define the ionized IGM as an arbitrary region in the $\varkappa-\delta$ plane rather than a region bounded by constant density and ionized fraction constraints. The thresholds for the two definitions of the IGM (corresponding to two pivot points) are shown with dotted and dashed lines in Figure \ref{fig:phaseplot}.

The proposed definition, in contrast to the choice of arbitrary fixed density and ionized fraction thresholds, uses the specific distribution of gas in the $\varkappa$-$\delta$ plane to define the ionized IGM - which makes the density and the ionization fraction cuts time dependent (and, potentially, different in different simulations). Therefore, the applied thresholds account for the evolving ionizing background and for structure formation.

Once we have defined the region in the $\delta-\varkappa$ phase space that we identify with the ionized IGM, we can calculate its clumping factor. Given a definition of the ionized IGM, we calculate the average squared ionized hydrogen density and divide it by the squared average ionized hydrogen density over all simulation cells that fall within the IGM definition,
\begin{equation}
C_{\rm IGM} = \dfrac{\sum_{i\in \rm IGM} V_i \sum_{i\in \rm IGM} V_i (1+\delta_i)^2 x_{\HII,\,i}^2}{\left(\sum_{i\in \rm IGM} V_i (1+\delta_i) x_{\HII,\,i}\right)^2},
\end{equation}
where $V_i$, $\delta_i$ and $x_{\HII,\,i}$ are the volume, the overdensity and the ionized fraction of the $i$-th cell correspondingly. 

In order to distinguish various definitions of the ionized IGM that we discuss above, we will use symbols $C_\Diamond$ and $C_\Box$ to label the definitions of the IGM based on the thresholds from Figure\ \ref{tab:thresh} and a symbol $C_{\delta,\,x_{\HII}}$ for the definition of the IGM from fixed thresholds in density and ionized fraction.

It is worth mentioning that all the information necessary for calculating the clumping factor is contained in the phase diagram. Hence, instead of iterating over cells in the simulation box, one can integrate over the area $S$ of the phase diagram, which is identified with the ionized IGM,
\begin{equation}
C_{\rm IGM} = \dfrac{\int_S N_M(1+\delta, \varkappa) \times (1+\delta)\; \mathrm{d}(1+\delta)\mathrm{d}\varkappa}{\int_S N_M(1+\delta, \varkappa) / (1+\delta)\; \mathrm{d}(1+\delta)\mathrm{d}\varkappa},
\end{equation}
where $N_M$ is mass weighted 2D histogram shown in Figure \ref{fig:phaseplot}. 

\begin{figure}
    \includegraphics[scale=1.0]{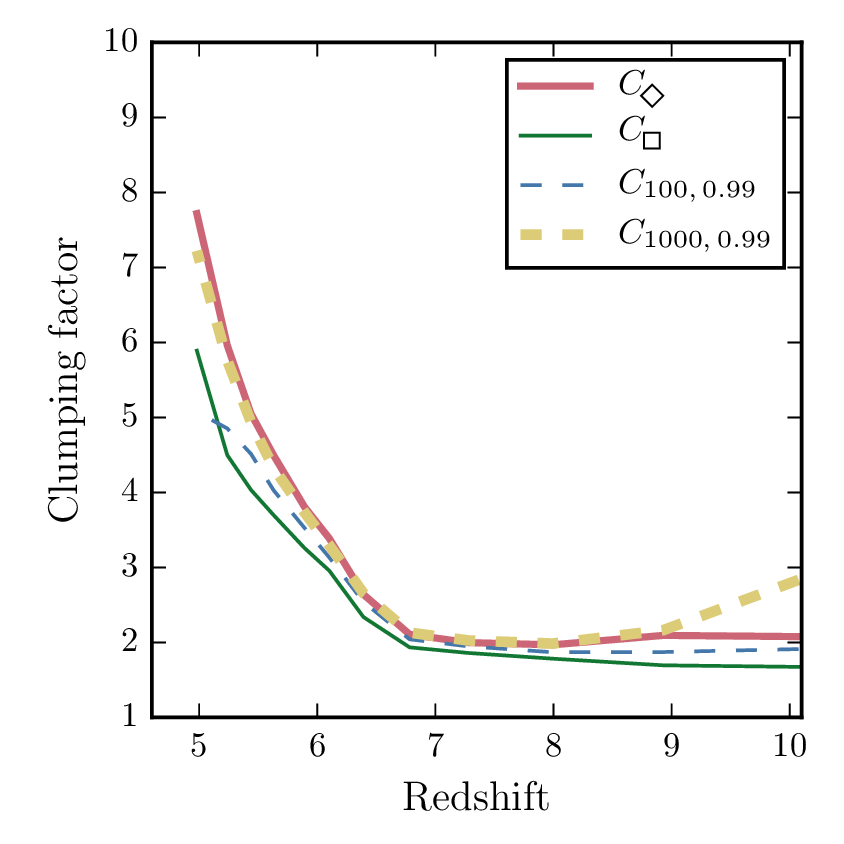}
    \caption{\label{fig:cf}Clumping factor of the IGM as a function of redshift. Solid lines represents our two adaptive definitions of the ionized IGM, while dashed lines show models with fixed thresholds.}
\end{figure}

The result of such calculations is presented in Figure \ref{fig:cf}. Notice, that the adaptive thresholds may change for a different reionization model, to reflect the actual onset of self-shielding in the IGM. By accident, for our particular model, $C_\Box$ and $C_{\delta<100,\;x_\HII>0.99}$ end up very similar.

Our two adopted definitions share the same ionization threshold, which does not change much with redshift in this particular model of reionization (see Figure \ref{tab:thresh}). On the other hand, the density threshold evolves significantly, and the one associated with $C_\Diamond$ is about two order of magnitude higher than the one used in $C_\Box$. Both definitions account for ionized IGM; however, $C_\Box$, in contrast to $C_\Diamond$, includes less volume associated with the proximity effect (see Figure \ref{fig:phaseplot}). This difference has a moderate influence on the global clumping factor at redshifts $z \gtrsim 6$ (see Figure \ref{fig:cf}), but reaches about 25\% at $z \sim 5$. Therefore, if $C_\Diamond$ definition is used, one needs to account for recombinations in the proximity zones by increasing proportionally the effective escape fractions from ionizing sources.

In contrast to previous works \citep{Shull2012,So2014,Jeeson-Daniel2014} where fixed density threshold is used for defining clumping factor, we propose to derive it directly from a simulation. In \cite{McQuinn2011} the connection between ionizing background and such a density threshold has been studied and a power law connection has been found. This result has been used to determine the density threshold in \cite{Finlator2012} study. Similar approach has been adopted in \cite{Emberson2013}, where the authors derive the critical density cut-off from their simulation and also find power law dependence. Our findings can be interpreted in a similar manner. The fact that the ionization fraction threshold (Figure \ref{tab:thresh}) does not change significantly means that both our markers lie on the same line in $\varkappa-\delta$ phase diagram at all redshifts. Therefore, taking into account Equation \ref{eq:varkappa1}, the density threshold $\delta$ has a power law dependence on average ionizing background $\Gamma$.

\subsection{Local variations of the clumping factor}
\label{sec:localclumpingfactor}

In addition to the temporal evolution of the global, averaged over the whole universe clumping factor, one may also be interested in its (in)homogeneity. We now explore spatial variations of the clumping factor at fixed redshift $z=5.7$, when the IGM is already highly ionized. In this section we use our definition of the clumping factor ($C_\Diamond$), however all general trends are expected to be similar for other definitions of the clumping factor as well.

In order to measure the spatial variations in the clumping factor in the simulations, we split the simulation box into cubic sub-boxes with sizes from $0.15\Mpch$ up to $20\Mpch$ (half the box size), and calculate the local clumping factor in each sub-box. We define the local clumping factor as:
\begin{equation}
 C_\mathrm{loc} = \dfrac{\langle n_i^2 \rangle_\mathrm{loc}}{\langle n_i \rangle_\mathrm{loc}^2} = 
 \dfrac{ \left\langle n_i^2 \right\rangle_\mathrm{loc} }{ \bar{n_i}_\mathrm{Universe}^2 \, (1+\bar\delta_\mathrm{loc})^2},
\end{equation} 
where $n_i$ is number density in ionized regions, $\langle\rangle_\mathrm{loc}$ is the average inside each cubic sub-box, and $\bar{n_i}_\mathrm{Universe}$ and ${\bar\delta}_\mathrm{loc}$ are the cosmic mean density of ionized regions and the average overdensity of the sub-box. Notice, that there is the $(1+\bar\delta_\mathrm{loc})^{-2}$ dependence in our definition that accounts for the given sub-box being over- or under-dense.

\begin{figure}    
    \includegraphics{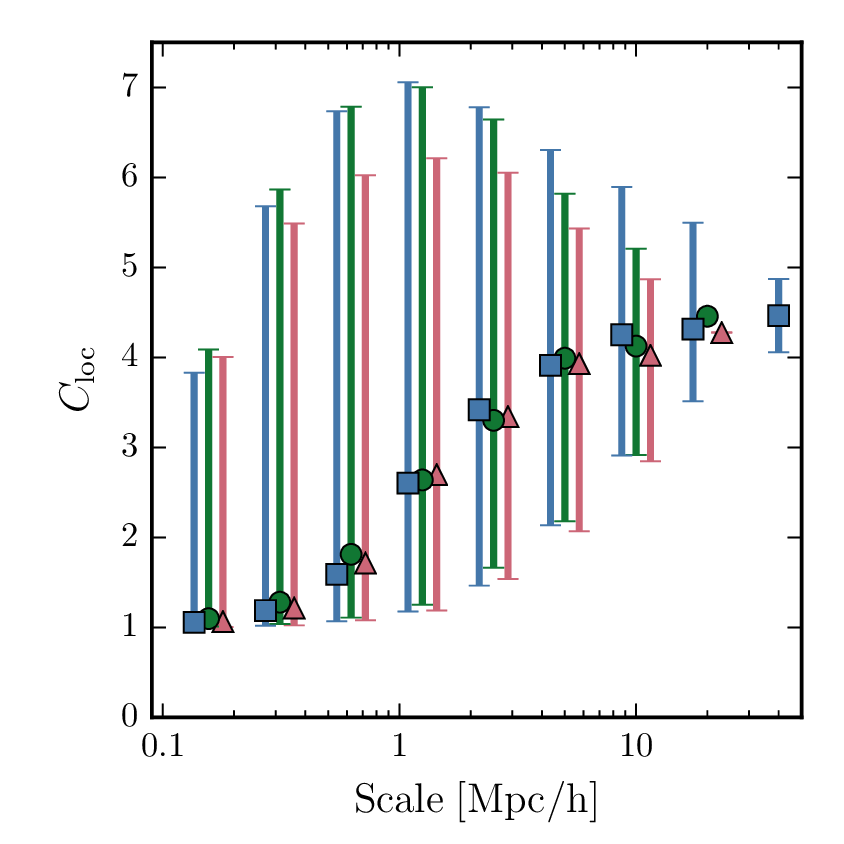}
    \caption{\label{fig:cfscaling} The distribution of local clumping factors at $z=5.7$ versus scale it is defined over. The marker shows the median value and errorbars trace its 1\%-99\% scatter range. Blue squares correspond to all three $40\Mpch$ simulations; the rightmost point and its errorbar correspond to the mean and the standard deviation of the clumping factor between the three independent realizations. Other two sets of points correspond to two $20\Mpch$ simulations with identical initial conditions but different resolutions. Simulation which correspond to the red triangle has the same resolution as our $40\Mpch$ runs, and the simulation shown with green circles has 8 times higher mass resolution and 2 times higher spatial resolution in the IGM.}
\end{figure}

The local clumping factor versus the spatial averaging scale is shown in Figure \ref{fig:cfscaling}. A couple of trends can be observed. First, the median value of the clumping factor approaches unity at small scales, and the scatter is lower in smaller sub-boxes. Reduced clumping factor implies that smaller sub-boxes contain less substructure; when the clumping factor approaches unity, the averaging scale becomes comparable to a smoothing scale, below which the density field is uniform. There are two candidates for this smoothing scale: physical -- the pressure smoothing scale, and numerical -- the finite resolution of the simulation. The first one is the scale over which the pressure of the photo-ionized gas erases baryonic fluctuations \citep{ng:gh98,Gnedin2003,Kulkarni:2015fga}. Therefore, the gas distribution in sub-boxes with sizes comparable to the pressure smoothing scale scale is more-or-less uniform, and, consequently, the clumping factor is close to unity. 

The exact value of the pressure smoothing scale depends on the variety of factors, such as thermal history and the degree of nonlinearity in the distribution of the particular subset of gas under consideration (namely, the gas that falls inside the cuts in Fig.\ \ref{fig:phaseplot}). In the post-reionization epoch ($z\sim 2-4$) the linear pressure smoothing scale (often called ``filtering scale'') is about $50-70\dim{kpc}$ \citep{Gnedin2003}; the nonlinear pressure smoothing scale, however, is somewhat larger, closer to $100\dim{kpc}$ at $z\sim3$ \citep{Kulkarni:2015fga}, and getting even larger (up to $200\dim{kpc}$) as the IGM temperature increases to close to $10^5\dim{K}$ (i.e.\ closer to reionization). The latter value is similar to (albeith still smaller than) the scale in Fig.\ \ref{fig:cfscaling} at which the local clumping factor approaches unity. The exact comparison between these two quantities, however, would require a numerically expensive focused study, whose value would be largely academic.

However, simulations do not have infinitely fine resolution, so it is also possible that the simulation does not resolve this scale, and observed smoothing is just a resolution effect. In order to exclude the latter possibility, we perform the same analysis on two $20\Mpch$ runs with identical initial conditions, but with different mass and spatial resolutions. The result is presented in Figure \ref{fig:cfscaling}. The clumping factor and its scatter are only insignificantly larger in a higher resolution simulation; hence the behavior of the clumping factor shown in Figure \ref{fig:cfscaling} is real and not a numerical artifact.

The second obvious trend in Figure \ref{fig:cfscaling} is that, as the averaging scale increases, the median gets closer to the global clumping factor, and the scatter also decreases. Even if the actual values of the clumping factor on large scales are affected by the finite size of the simulation box, the qualitative behavior is as expected, since at the largest scales the universe is approaching homogeneity. 

The scatter of the clumping factor peaks at the intermediate scale of a few $\Mpch$. This scale is in the same order-of-magnitude range as several physical scales in the problem (galaxy clustering scale, typical size of ionized bubbles, the photon mean free path due to LLS, etc), so the reason behind the increase of scatter would be virtually impossible to isolate.

\subsection{Local clumping and density correlation}
\label{subsec:localclump}

\begin{figure}
    \begin{center}
    \includegraphics[scale=0.495]{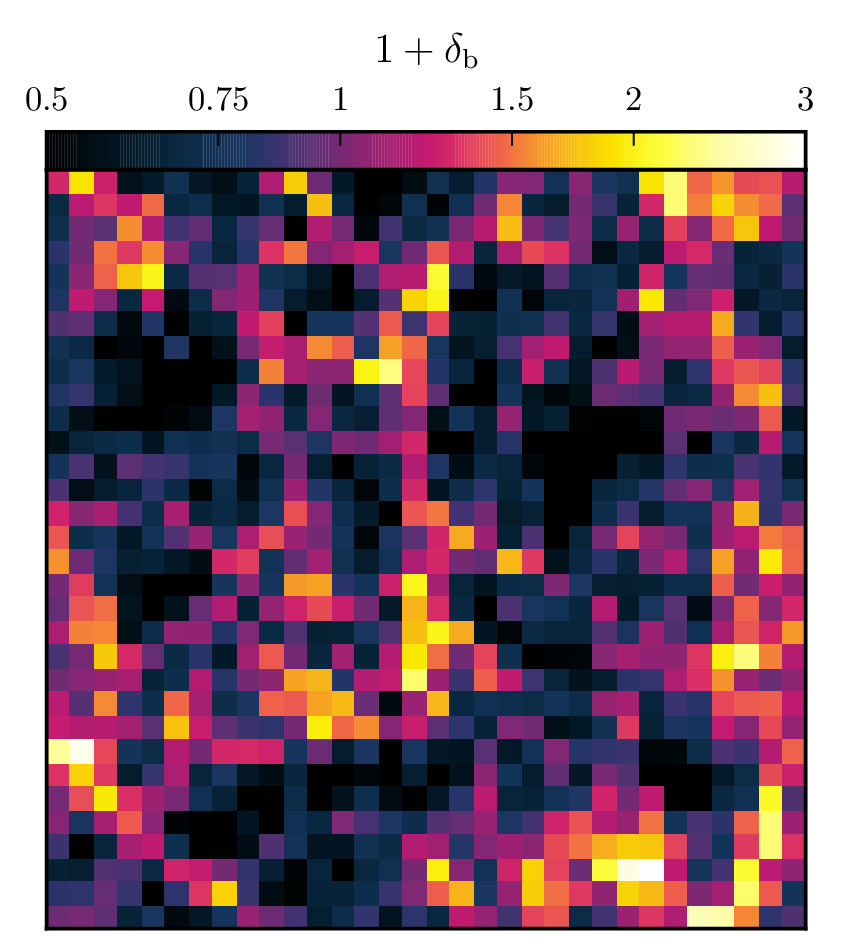}
    \includegraphics[scale=0.495]{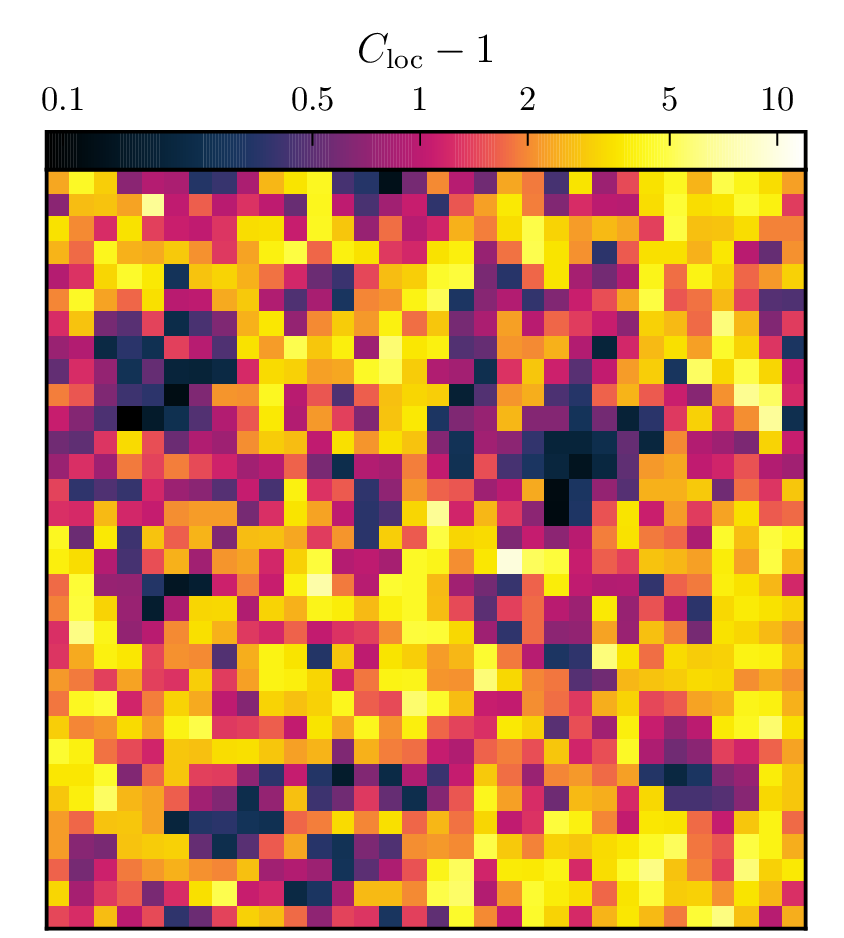}
    \end{center}
    \caption{\label{fig:cfdistr} The slice of $40\Mpch$ simulation at redshift 5.7. Colors represents the baryon density field (left panel), the local clumping factor (right panel). Both fields are defined on $1.25\Mpch$ scale.}
\end{figure}

The scatter in the local clumping factor from Figure \ref{fig:cfscaling} is not necessarily random. It can correlate with several properties of sub-boxes, of which the mean density is the primary candidate. In Figure \ref{fig:cfdistr} slices of density and the clumping factor in $1.25h^{-1}\dim{Mpc}$ sub-boxes are presented. There is an obvious correlation between them. It motivates us to study this dependence in more detail. 

\begin{figure}
    \includegraphics{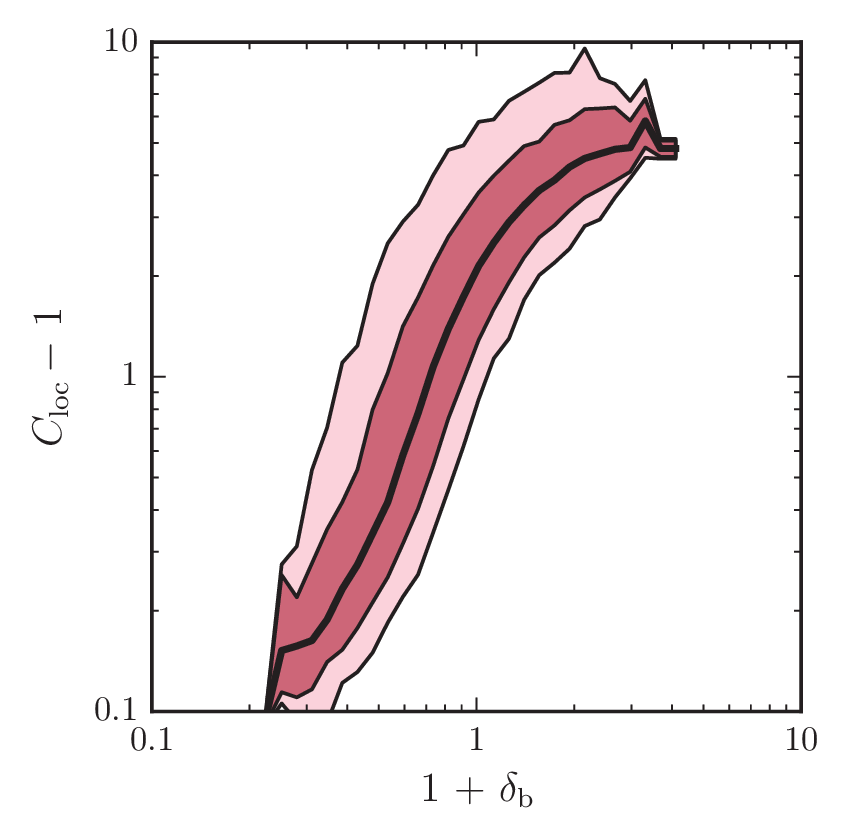}
    \caption{\label{fig:cfmap}Local clumping factor versus average local density at $z=5.7$ defined at scale of $1.25\Mpch$. Black solid line represents median and contours show 10\%-90\% and 1\%-99\% intervals.}
\end{figure}

We group sub-boxes by density and measure the distributions of clumping factors within sub-boxes of the same density. In Figure \ref{fig:cfmap} we show the scatter of local clumping factor as a function of the mean density in $1.25h^{-1}\dim{Mpc}$ sub-boxes. The correlation is apparent, but the scatter of the clumping factor at fixed density is still over a factor of 3-5. Nevertheless, this dependence can be used for sub-grid modeling of clumping factor in low resolution simulations.

By definition, the clumping factor is nothing else but the second moment (variance) of the probability distribution function (PDF) of density. Therefore we take a look at PDFs within sub-boxes with a given mean density, in order to explore what features in the PDF lead to the increased clumping factor in denser regions. 

\begin{figure*}
    \includegraphics{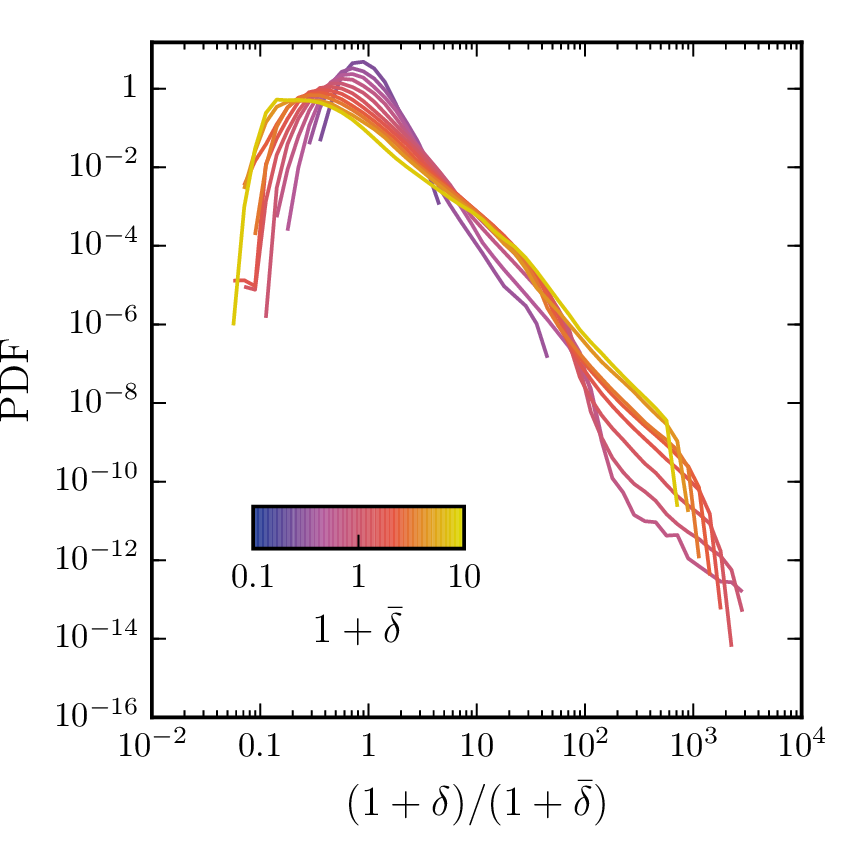}
    \includegraphics{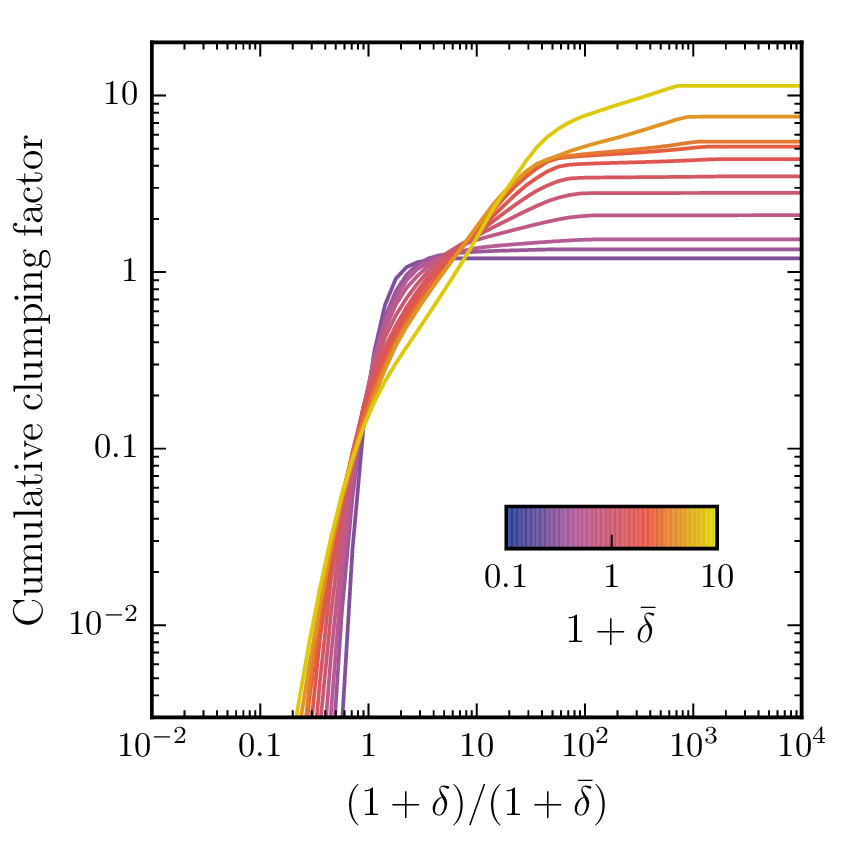}
    \caption{\label{fig:pdf}Left panel: probability distribution functions (PDF) of ionized baryon density in $1.25\Mpch$ sub-boxes of given mean density (color coded) \textbf{at $z=5.7$}. Right panel: the cumulative clumping factor as a function of density in the same sub-boxes (i.e., the variance (second moment) of the PDFs in the left panel as a function of the maximum density of integration).}
\end{figure*}

These PDFs are presented in the left panel of Figure \ref{fig:pdf}. The observed shapes correspond to neither normal nor log-normal distributions, and reveal a power law slope at high densities, followed by a partial break at densities where the self-shielding sets in (as a reminder, we only consider densities of the ionized gas, since only they contribute to the clumping factor, Equation \ref{eq:cf}). In sub-boxes of higher mean densities the self-shielding sets in at proportionally higher densities, so that when the densities in each sub-box are scaled by the mean sub-box density, the self-shielding threshold remains roughly constant, $(1+\delta)/(1+\bar\delta)\approx 100$ (at this redshift and for this reionization model). 


From the shapes of PDFs in the left panel of Figure \ref{fig:pdf} it is not immediately clear which range of densities contributes most to the clumping factor. Therefore, we show in the right panel of Figure \ref{fig:pdf} the cumulative clumping factor as a function of the maximum density for the PDFs from the left panel. As one can see, almost all of the contribution to the clumping factor comes from modest densities, $(1+\delta)/(1+\bar\delta)\la 10-100$, well below the self-shielding threshold. Hence, the dominant contribution to the clumping factor comes form the densities around the peak of the PDF, and not from the high density tail. Hence, the increase of the clumping factor with the density is not unexpected: denser regions, being analogous to denser universes, are more evolved and, hence, have a wider density distribution.

\section{Conclusions}
\label{sec:concl}

Spontaneous recombination of ionized hydrogen takes away ionizing photons, and, hence, is an important physical process during and after cosmic reionization. Even though recombination is the same physical process no matter where it occurs, it is customarily quantified in analytical studies and in not-fully-self-consistent simulations with three separate quantities: the escape fraction, the IGM clumping factor, and the maximum mean free path. 

Using fully self-consistent numerical simulations of cosmic reionization, we explore whether such a separation is physically motivated and robust. To that end, we use a convenient physical quantity, the ``ionization state indicator'' $\varkappa$ and a $\varkappa-\delta$ ``phase diagram'', to introduce a well-motivated and approximately robust definition of the IGM, and to compute its clumping factor. In comparing to previous work, we find that our physical definition of the IGM is reasonably well approximated by simple fixed thresholds in ionized fraction ($x_{\HII}>0.99$) and density ($\delta\ga10^2-10^3$).

The largest ambiguity in the definition of the ionized IGM comes from the unvirialized regions around galaxies that are over-ionized by the local enhancement in the radiation field (``proximity zones''). That inherent ambiguity imposes a ``systematic error'' on the value of the clumping factor of about 20\% (somewhat smaller during reionization but increasing to $\approx$25\% at lower redshifts). 

The fact that self-shielded neutral regions separate cleanly from the ionized IGM allows one to account for them separately. In ionized IGM recombinations are proportional to density squared, and therefore the clumping factor is a convenient descriptor. Recombinations on the surface of self-shielded regions only compensates for the ionizations from external radiation, and in that case the maximum mean free path of ionizing photons is an appropriate quantity.

The last regime of ``three-regime'' approach is galactic interiors. Even though our simulations do not resolve internal structure of galaxies, the compactness of galaxies and their mutual separation allow to consider them as isolated systems. Therefore, characterizing the escape of photons with a single number (the escape fraction) and neglecting angular inhomogeneity may be sufficient for many studies.

We also explore the scale-dependence of the clumping factor over the range of scales, faithfully represented in our simulations. We find that the clumping factor computed in sub-boxes of a given size correlates strongly, but not perfectly, with the mean density in such sub-boxes. This correlation is driven by the increase in the PDF width in denser sub-boxes (and not by their high density tails), which, being analogous to denser universes, are more evolved and, hence, have a wider density distribution.

Nevertheless, the correlation between the local clumping factor and the mean density over the scale it is computed is not perfect, and other factors introduce significant (factor of 3-5) scatter in the relation. In principle, numerical simulations would allow us to further explore that additional dependence; however, we do not engage in such study in this work, as its practical need is not presently clear.


\acknowledgements 
Fermilab is operated by Fermi Research Alliance, LLC, under Contract
No.~DE-AC02-07CH11359 with the United States Department of Energy.
This work was also supported in part by the NSF grant AST-1211190 and by the NASA grant NNX-09AJ54G. This work made extensive use of the NASA Astrophysics Data System and {\tt arXiv.org} preprint server. This work was done with significant usage of YT package \citep{Turk2011}.

\bibliographystyle{apj}
\bibliography{bibiki,ng-bibs/self,ng-bibs/newrei}

\begin{thebibliography}{29}
\expandafter\ifx\csname natexlab\endcsname\relax\def\natexlab#1{#1}\fi

\bibitem[{Emberson {et~al.}(2013)Emberson, Thomas, \& Alvarez}]{Emberson2013}
Emberson, J., Thomas, R.~M., \& Alvarez, M.~A. 2013, Astrophys.J., 763, 146

\bibitem[{{Finlator} {et~al.}(2012){Finlator}, {Oh}, {{\"O}zel}, \&
  {Dav{\'e}}}]{Finlator2012}
{Finlator}, K., {Oh}, S.~P., {{\"O}zel}, F., \& {Dav{\'e}}, R. 2012, \mnras,
  427, 2464

\bibitem[{Gnedin(2014)}]{Gnedin2014a}
Gnedin, N.~Y. 2014, Astrophys.J., 793, 29

\bibitem[{Gnedin {et~al.}(2003)Gnedin, Baker, Bethell, Drosback, Harford,
  {et~al.}}]{Gnedin2003}
Gnedin, N.~Y., Baker, E.~J., Bethell, T.~J., {et~al.} 2003, Astrophys.J., 583,
  525

\bibitem[{{Gnedin} \& {Hui}(1998)}]{ng:gh98}
{Gnedin}, N.~Y., \& {Hui}, L. 1998, \mnras, 296, 44

\bibitem[{Gnedin \& Kaurov(2014)}]{Gnedin2014b}
Gnedin, N.~Y., \& Kaurov, A.~A. 2014, Astrophys.J., 793, 30

\bibitem[{{Gnedin} {et~al.}(2008){Gnedin}, {Kravtsov}, \& {Chen}}]{Gnedin2008a}
{Gnedin}, N.~Y., {Kravtsov}, A.~V., \& {Chen}, H.-W. 2008, \apj, 672, 765

\bibitem[{{Gnedin} {et~al.}(2011){Gnedin}, {Kravtsov}, \& {Rudd}}]{gkr11}
{Gnedin}, N.~Y., {Kravtsov}, A.~V., \& {Rudd}, D.~H. 2011, \apjs, 194, 46

\bibitem[{Gnedin \& Ostriker(1997)}]{Gnedin1997a}
Gnedin, N.~Y., \& Ostriker, J.~P. 1997, Astrophys.J., 486, 581

\bibitem[{{Hutter} {et~al.}(2014){Hutter}, {Dayal}, {Partl}, \&
  {M{\"u}ller}}]{newrei:hdp14}
{Hutter}, A., {Dayal}, P., {Partl}, A.~M., \& {M{\"u}ller}, V. 2014, \mnras,
  441, 2861

\bibitem[{Iliev {et~al.}(2005)Iliev, Scannapieco, \& Shapiro}]{Iliev2005}
Iliev, I.~T., Scannapieco, E., \& Shapiro, P.~R. 2005, Astrophys.J., 624, 491

\bibitem[{Jeeson-Daniel {et~al.}(2014)Jeeson-Daniel, Ciardi, \&
  Graziani}]{Jeeson-Daniel2014}
Jeeson-Daniel, A., Ciardi, B., \& Graziani, L. 2014, Mon.Not.Roy.Astron.Soc.,
  443, 2722

\bibitem[{{Kaurov} \& {Gnedin}(2013)}]{Kaurov2013a}
{Kaurov}, A.~A., \& {Gnedin}, N.~Y. 2013, \apj, 771, 35

\bibitem[{Kaurov \& Gnedin(2014)}]{Kaurov2014}
Kaurov, A.~A., \& Gnedin, N.~Y. 2014, Astrophys.J., 787, 146

\bibitem[{Kohler {et~al.}(2007)Kohler, Gnedin, \& Hamilton}]{Kohler2007}
Kohler, K., Gnedin, N.~Y., \& Hamilton, A.~J. 2007, Astrophys.J., 657, 15

\bibitem[{{Kuhlen} \& {Faucher-Gigu{\`e}re}(2012)}]{Kuhlen2012}
{Kuhlen}, M., \& {Faucher-Gigu{\`e}re}, C.-A. 2012, \mnras, 423, 862

\bibitem[{Kulkarni {et~al.}(2015)Kulkarni, Hennawi, OÃ±orbe, Rorai, \&
  Springel}]{Kulkarni:2015fga}
Kulkarni, G., Hennawi, J.~F., OÃ±orbe, J., Rorai, A., \& Springel, V. 2015

\bibitem[{McQuinn {et~al.}(2007)McQuinn, Lidz, Zahn, Dutta, Hernquist,
  {et~al.}}]{McQuinn2007}
McQuinn, M., Lidz, A., Zahn, O., {et~al.} 2007, Mon.Not.Roy.Astron.Soc., 377,
  1043

\bibitem[{McQuinn {et~al.}(2011)McQuinn, Oh, \&
  Faucher-Gigu{\`e}re}]{McQuinn2011}
McQuinn, M., Oh, S.~P., \& Faucher-Gigu{\`e}re, C.-A. 2011, The Astrophysical
  Journal, 743, 82

\bibitem[{{Norman} {et~al.}(2013){Norman}, {Reynolds}, {So}, \&
  {Harkness}}]{newrei:nrs14}
{Norman}, M.~L., {Reynolds}, D.~R., {So}, G.~C., \& {Harkness}, R.~P. 2013,
  ArXiv e-prints

\bibitem[{Pawlik {et~al.}(2008)Pawlik, Schaye, \& van
  Scherpenzeel}]{Pawlik2009}
Pawlik, A.~H., Schaye, J., \& van Scherpenzeel, E. 2008, \mnras, 394, 1812

\bibitem[{{Rai{\v c}evi{\'c}} \& {Theuns}(2011)}]{Raivcevic2011}
{Rai{\v c}evi{\'c}}, M., \& {Theuns}, T. 2011, \mnras, 412, L16

\bibitem[{{Shull} {et~al.}(2012){Shull}, {Harness}, {Trenti}, \&
  {Smith}}]{Shull2012}
{Shull}, J.~M., {Harness}, A., {Trenti}, M., \& {Smith}, B.~D. 2012, \apj, 747,
  100

\bibitem[{So {et~al.}(2013)So, Norman, Reynolds, \& Harkness}]{So2014}
So, G.~C., Norman, M.~L., Reynolds, D.~R., \& Harkness, R.~P. 2013, \apj, 789,
  149

\bibitem[{{So} {et~al.}(2014){So}, {Norman}, {Reynolds}, \&
  {Wise}}]{newrei:snr14}
{So}, G.~C., {Norman}, M.~L., {Reynolds}, D.~R., \& {Wise}, J.~H. 2014, \apj,
  789, 149

\bibitem[{{Sobacchi} \& {Mesinger}(2014)}]{Sobacchi2014}
{Sobacchi}, E., \& {Mesinger}, A. 2014, \mnras, 440, 1662

\bibitem[{{Songaila} \& {Cowie}(2010)}]{Songaila2010}
{Songaila}, A., \& {Cowie}, L.~L. 2010, \apj, 721, 1448

\bibitem[{Trac \& Cen(2007)}]{Trac2007}
Trac, H., \& Cen, R. 2007, Astrophys.J., 671, 1

\bibitem[{{Turk} {et~al.}(2011){Turk}, {Smith}, {Oishi}, {Skory}, {Skillman},
  {Abel}, \& {Norman}}]{Turk2011}
{Turk}, M.~J., {Smith}, B.~D., {Oishi}, J.~S., {et~al.} 2011, \apjs, 192, 9

\end{thebibliography}


\end{document}